\documentclass[aps,pre,reprint,showpacs]{revtex4-1}
\usepackage[latin1]{inputenc}
\usepackage{graphicx}
\usepackage{amsmath,amssymb,amstext}

\makeatletter
\def\imod#1{\allowbreak\mkern10mu({\operator@font mod}\,\,#1)}
\makeatother

\DeclareMathOperator*{\rank}{rank}

\DeclareMathOperator*{\nRe}{Re}
\DeclareMathOperator*{\nIm}{Im}

\begin{document}

\date{\today}

\title{Cluster and group synchronization in delay-coupled networks}

\author{Thomas Dahms}
\author{Judith Lehnert}
\author{Eckehard Sch\"{o}ll}
\email{schoell@physik.tu-berlin.de}

\affiliation{Institut f{\"u}r Theoretische Physik, Technische
  Universit{\"a}t Berlin, 10623 Berlin, Germany}

\begin{abstract}
  We investigate the stability of synchronized states in delay-coupled
  networks where synchronization takes place in groups of different
  local dynamics or in cluster states in networks with identical local
  dynamics. Using a master stability approach, we find that the master
  stability function shows a discrete rotational symmetry depending on
  the number of groups. The coupling matrices that permit solutions
  on group or cluster synchronization manifolds show a very similar
  symmetry in their eigenvalue spectrum, which helps to simplify the
  evaluation of the master stability function. Our theory allows for
  the characterization of stability of different patterns of
  synchronized dynamics in networks with multiple delay times,
  multiple coupling functions, but also with multiple kinds of local
  dynamics in the networks' nodes.  We illustrate our results by
  calculating stability in the example of delay-coupled semiconductor
  lasers and in a model for neuronal spiking dynamics.
\end{abstract}


\maketitle

\section{Introduction}

The scientific field of synchronization in coupled systems has evolved
rapidly in the last decades
\cite{STR93,ROS96,BRO97,PIK01,BOC02,MOS02,BAL09}. Complete or
isochronous synchronization of coupled chaotic units
\cite{PEC90,KOC95,PEC97,KIN09} as well as of time-periodic systems has
been extensively studied \cite{EAR03,DHU08,SET08}. In general, more
complicated synchronization patterns may be observed including
cluster, group, and sublattice synchronization
\cite{KES07,KES08,KAN11,KAN11a,AVI12}. Cluster synchronization, where
certain clusters inside the network show isochronous synchronization,
will be investigated in this paper. Additionally, we describe the
stability of group synchronization, i.e., a generalization of cluster
synchronization where the local dynamics of the nodes in each group
differ.

The characterization of stability of isochronous synchronization has
been widely studied, and the ground-breaking work by Pecora and
Carroll \cite{PEC98} which allows for a separation of network topology
and local dynamics of the nodes was recently also applied to networks
with delays in the links \cite{KIN09,ENG10,FLU10b,HEI11,FLU11b}. Such
delay times can greatly change the synchronization properties and
appear in many natural coupled systems. For example, in optical
applications delay times arise from the finite speed of light and in
neuronal networks delays play a role due to finite distances between
interacting neurons, but also due to processing lags in the neurons.

For group and cluster synchronization, attempts have been made to
treat stability within a master stability approach. Sorrentino and Ott
\cite{SOR07} considered two groups of nodes governed by different
local dynamics. In the present paper, we show how this can be
generalized to a higher number of groups and what restrictions for the
topology of the network arise. Moreover, our framework allows us to
have multiple delay times in the network. Making use of a separation of
the topologies into multiple coupling matrices we can lift the
restriction that no coupling may exist inside groups or clusters,
i.e., a restriction to multipartite topologies. This makes our theory
accessible for a wide range of topologies.

After introducing the notion of cluster and group dynamics in
Sec.~\ref{sec:cluster-dynamics}, we derive the master stability
function and show the restrictions that arise upon the topology in
Sec.~\ref{sec:stab-clust-synchr}. In Sec.~\ref{sec:symm-mast-stab} we
investigate the symmetries that group and cluster synchronization
impose on the master stability function. In Sec.~\ref{sec:laser}, we
demonstrate this symmetry for networks of delay-coupled
lasers. Multiple coupling matrices are introduced in
Sec.~\ref{sec:beyond-one-block}, where we use a hierarchical network structure
as an example. The effect of different delay times is shown for the
example of neuronal networks in
Sec.~\ref{sec:exampl-neur-netw}. Finally, we conclude with
Sec.~\ref{sec:conclusion}.

\section{Cluster and group dynamics}\label{sec:cluster-dynamics}

In a network consisting of $N$ identical nodes, we refer to cluster
synchronization as a state where clusters of nodes exist that show
isochronous synchronization internally, but synchronization between
these cluster does not occur, or is of non-isochronous type, i.e.,
there may be a phase lag between clusters \cite{CHO09,SEL12}.

Group synchronization describes a similar state of synchrony, but the
node dynamics -- determined by the functional form of the local dynamics 
-- differs from cluster to cluster. We refer to these clusters as {\em groups}.
As cluster synchronization is a special case of group
synchronization, we use the more general notion of groups in the
following.

Assume the number of groups to be $M$ where $k=1,\ldots,M$ numbers
the individual groups. The dynamical variables of the nodes in each
group are then given by $\mathbf{x}_i^{(k)} \in \mathbb{R}^{d_k}$ with
$i=1,\ldots,N_k$, where $N_k$ denotes the number of nodes in the
$k$-th group. The dimension $d_k$ of the $\mathbf{x}_i^{(k)}$
is given by the particular node model, e.g., the complex
Hopf normal-form (Stuart-Landau) oscillator \cite{CHO09}, the two-dimensional
FitzHugh-Nagumo model \cite{LEH11}, or the three-dimensional
Lang-Kobayashi equations \cite{FLU09}.

In general the dimension $d_k$ of the nodes $\mathbf{x}_i^{(k)}$ may be
different for each group $k$. Consequently, also the local dynamics
$\mathbf{F}^{(k)}(\mathbf{x}_i^{(k)})$ can be different for each
group, but must be identical for all nodes $i=1,\ldots,N_k$ in a
given group $k$. For example, consider a network of neurons, where
one group contains inhibitory neurons and another group contains
excitatory ones. The local dynamics will be different for each group,
and depending on the model used to describe both types of neurons
also the dimension of the node dynamics may be different.

Let $\sigma^{(k)}$ be the coupling strength for the coupling from the
$(k-1)$-th to the $k$-th group. In the same sense, let
$\mathbf{A}^{(k)}$ be an $N_{k-1} \times N_k$ coupling matrix, such
that its entries $\{A^{(k)}_{ij}\}$ represent the coupling of node $j$
(which is in the $(k-1)$-th group) to node $i$ (which is in the $k$-th
group).  By this construction we obtain a multipartite topology in
which one cluster has incoming links from only one neighbor while
having outgoing links to another one. The stability analysis performed
in this Section works for these topologies; but we will lift this
restriction by allowing multiple coupling matrices in
Sec.~\ref{sec:beyond-one-block}. Without loss of generality we assume
the row sums of the coupling matrices $\mathbf{A}^{(k)}$ to be unity,
which corresponds to the condition of unity or constant row sum needed
in the special case of complete isochronous synchronization
\cite{PEC98}. If a coupling matrix $\mathbf{A}^{(k)}$ has arbitrary
non-zero but constant row sum, unity row sum can easily be obtained by
rescaling the corresponding coupling strength $\sigma^{(k)}$.

As coupling schemes $\mathbf{H}^{(k)}$ we introduce $d_{k-1} \times d_k$
matrices, given that $d_{k-1}$ and $d_k$ are the dimensions of 
$\mathbf{x}_i^{(k-1)}$ and $\mathbf{x}_i^{(k)}$, i.e., the
dimensions of the local dynamics in the $(k-1)$-th and $k$-th group,
respectively. Note that, as a generalization, nonlinear coupling functions
$\mathbf{H}^{(k)}: \mathbb{R}^{d_{k-1}}\rightarrow\mathbb{R}^{d_k}$
may also be used instead of matrices \cite{PEC98,SOR07}.

\begin{figure}
  \includegraphics[width=\linewidth]{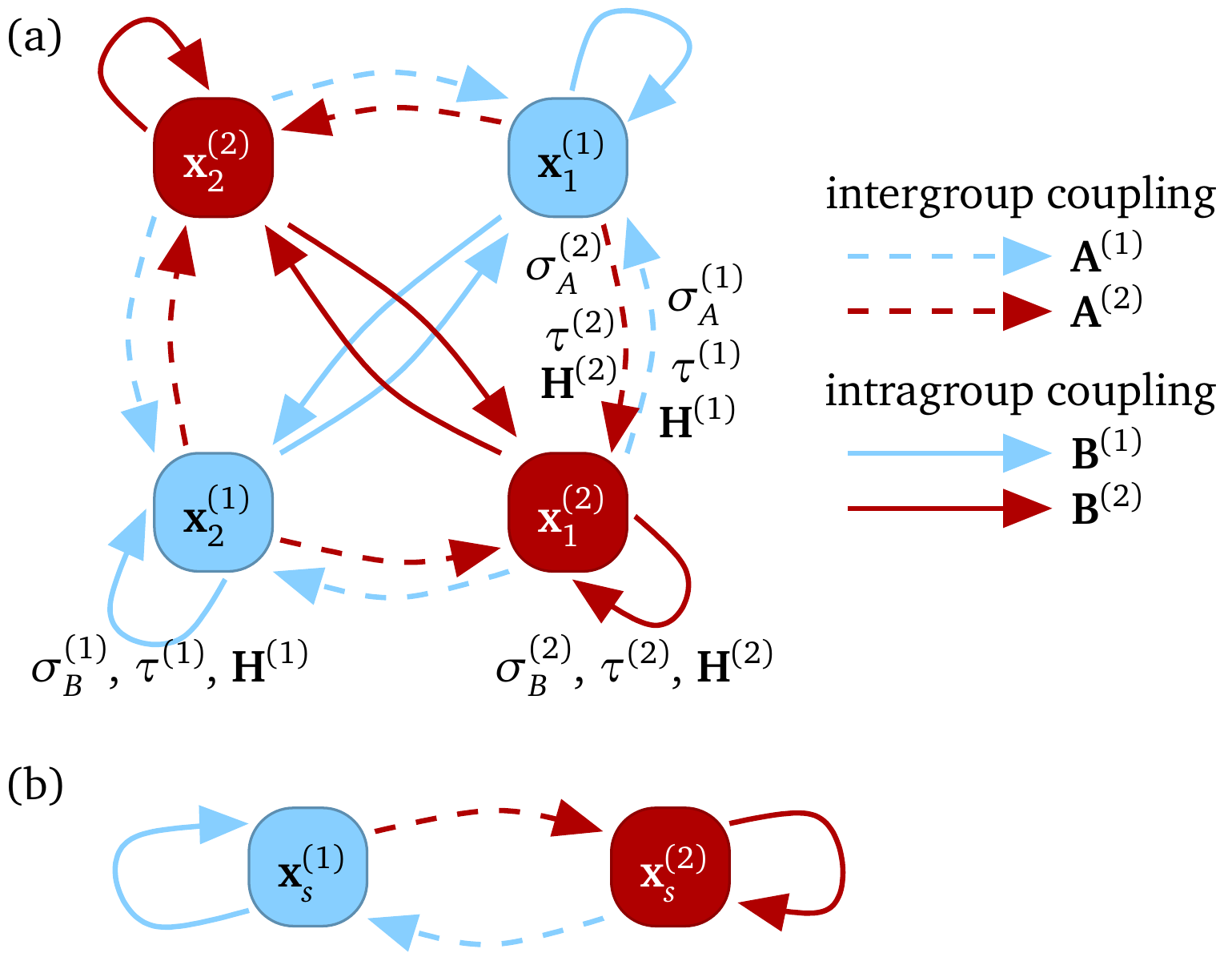}
  \caption{(Color online) (a) Schematic diagram of two groups
    visualizing parameters and dynamical variables as in
    Eq.~\eqref{eq:group_diff_eqn} for multipartite topologies (dashed
    arrows only, $\sigma_A^{(k)}\equiv\sigma^{(k)}$) and as in
    Eq.~\eqref{eq:79} for multiple coupling matrices (dashed and solid
    arrows). (b) The corresponding synchronization manifold according
    to Eqs.~\eqref{eq:group_sync_manifold} and \eqref{eq:80}.}
  \label{fig:schema}
\end{figure}
Finally, we allow the coupling delays $\tau^{(k)}$ to be different for
any pair $(k,k-1)$ of groups being connected. A schematic diagram of
the variables and matrices is shown in Fig.~\ref{fig:schema}(a). At
this point we consider only multipartite topologies, i.e., only the
dashed arrows in the Figure.

The dynamics of any single node in the network can then be described 
by the differential equation
\begin{equation}
  \dot{\mathbf{x}}_i^{(k)} = \mathbf{F}^{(k)}(\mathbf{x}_i^{(k)})
  + \sigma^{(k)} \sum_{j=1}^{N_{k-1}} A^{(k)}_{ij}
  \mathbf{H}^{(k)}\mathbf{x}_j^{(k-1)}(t-\tau^{(k)}).
  \label{eq:group_diff_eqn}
\end{equation}
for $i,j=1,\ldots,N_k$, $k=1,\ldots,M$.
This type of coupling is applicable for optical
systems \cite{LAN80b} and electronic circuits. In other cases, 
for instance neural dynamics, a diffusive-like coupling term of the form
$\sum_{j=1}^{N_{k-1}} A^{(k)}_{ij} \mathbf{H}^{(k)}
[\mathbf{x}_j^{(k-1)}(t-\tau^{(k)}) - \mathbf{x}_i^{(k)}(t)]$ is 
used. Both forms are equivalent since the local dynamics
can be transformed by
$\mathbf{F}^{(k)}(\mathbf{x}_i^{(k)}) \rightarrow
\mathbf{F}^{(k)}(\mathbf{x}_i^{(k)}) - \sigma^{(k)}\mathbf{H}^{(k)}
\mathbf{x}_i^{(k)}$. In the following, we will use the
form of Eq.~\eqref{eq:group_diff_eqn}.

The group synchronization manifold is then given by
\begin{equation}
  \label{eq:group_sync_manifold}
\dot{\mathbf{x}}_s^{(k)} = \mathbf{F}^{(k)}(\mathbf{x}_s^{(k)}) + \sigma^{(k)}
\mathbf{H}^{(k)}\mathbf{x}_s^{(k-1)}(t-\tau^{(k)}),
\end{equation}
which follows by inserting $\mathbf{x}_i^{(k)}=\mathbf{x}_j^{(k)}
\equiv \mathbf{x}_s^{(k)}$ into Eq.~\eqref{eq:group_diff_eqn}
($\forall i,j=1,\ldots,N_k$, $\forall k=1,\ldots,M$). For the example
of two groups, Fig.~\ref{fig:schema}(b) illustrates the
synchronization manifold, where Eq.~\eqref{eq:group_sync_manifold}
corresponds to the dashed arrows only.

Note that each group $k$ may exhibit different synchronous dynamics. Even
if the functions $\mathbf{F}^{(k)}$, the coupling matrices
$\mathbf{H}^{(k)}$, and the delay times $\tau^{(k)}$ are identical for
each group, different initial conditions can lead to
different dynamics.

\section{Stability of group synchronization}\label{sec:stab-clust-synchr}

In order to investigate the stability of the synchronous state, we
linearize Eq.~(\ref{eq:group_diff_eqn}) around the group synchronization manifold
$\mathbf{x}_s^{(k)}$ $(k=1,\ldots,M)$:
\begin{eqnarray}
  \delta\dot{\mathbf{x}}_i^{(k)} &=&
  D\mathbf{F}^{(k)}(\mathbf{x}_s^{(k)})\delta\mathbf{x}_i^{(k)}
  \nonumber \\
  &&+ \sigma^{(k)} \sum_{j=1}^{N_{k-1}} A^{(k)}_{ij} \mathbf{H}^{(k)}\delta\mathbf{x}_j^{(k-1)}(t-\tau^{(k)}).
  \label{eq:group_var_eqn}
\end{eqnarray}

Now assume that for each group $k=1,\ldots,M$ each of the $N_k$ solutions of
Eq.~\eqref{eq:group_var_eqn} can be written in the form
\begin{equation}
  \label{eq:group_linear_combi}
  \delta\mathbf{x}_i^{(k)} = c_i^{(k)} \delta\bar{\mathbf{x}}^{(k)},
\end{equation}
with time-independent scalars $c_i^{(k)}\in \mathbb{C}$. We show that the vectors
formed from the possible combinations of the
$c_{i_1}^{(1)},\ldots,c_{i_M}^{(M)}$ $(i_1=1,\ldots,N_1; \ldots;
i_M=1,\ldots,N_M)$ span a space of dimension $\sum_{k=1}^M N_k$, thus
the form~\eqref{eq:group_linear_combi} yields all solutions of
Eq.~\eqref{eq:group_var_eqn} as linear combinations. Using the
form~\eqref{eq:group_linear_combi}, Eq.~\eqref{eq:group_var_eqn}
becomes
\begin{eqnarray}
c_i^{(k)}\delta\dot{\bar{\mathbf{x}}}^{(k)} &=&
c_i^{(k)}D\mathbf{F}^{(k)}(\mathbf{x}_s^{(k)})\delta\bar{\mathbf{x}}^{(k)}
\label{eq:group_var_with_linear_combi} \\
&&+
\left( \sum_{j=1}^{N_{k-1}} A^{(k)}_{ij} c_j^{(k-1)} \right)
\sigma^{(k)} \mathbf{H}^{(k)}\delta\bar{\mathbf{x}}^{(k-1)}(t-\tau^{(k)}).
  \nonumber
\end{eqnarray}
Eq.~\eqref{eq:group_var_with_linear_combi} can be rewritten as
\begin{eqnarray}
\delta\dot{\bar{\mathbf{x}}}^{(k)} &=&
D\mathbf{F}^{(k)}(\mathbf{x}_s^{(k)})\delta\bar{\mathbf{x}}^{(k)}
\nonumber \\
&&+
C^{(k)} \sigma^{(k)} \mathbf{H}^{(k)}\delta\bar{\mathbf{x}}^{(k-1)}(t-\tau^{(k)}),
  \label{eq:group_var_with_linear_combi_cond}
\end{eqnarray}
assuming that
\begin{equation}
  \label{eq:cond_one_contribution}
  C^{(k)} = \frac{1}{c_i^{(k)}} \sum_{j=1}^{N_{k-1}} A^{(k)}_{ij}
  c_j^{(k-1)}
\end{equation}
is independent of $i=1,\ldots,N_k$. This is the case if a set of $M$
linearly independent vectors $\mathbf{c}^{(k)}=(c_1^{(k)}, c_2^{(k)},
\ldots, c_{N_k}^{(k)})$, $k=1,\ldots,M$, can be found, which we show
in the following. Using these vectors $\mathbf{c}^{(k)}$, the
conditions \eqref{eq:cond_one_contribution} can be written as
\begin{equation}
  \label{eq:cond_matrix_form} 
  \mathbf{A}^{(k)}\mathbf{c}^{(k-1)} = C^{(k)}\mathbf{c}^{(k)}.
\end{equation}
One particular solution $\mathbf{c}^{(k-1)},\mathbf{c}^{(k)}$ of
Eq.~\eqref{eq:cond_matrix_form} (and equivalently of
Eq.~\eqref{eq:group_var_with_linear_combi_cond}) is obtained when
setting $C^{(1)}=C^{(2)}=\ldots=C^{(M)}=\gamma$:
\begin{equation}
  \label{eq:cond_matrix_form_special}
  \mathbf{A}^{(k)}\mathbf{c}_0^{(k-1)} = \gamma\mathbf{c}_0^{(k)}.
\end{equation}
Introducing $M-1$ rescaling factors $z_1,\ldots,z_{M-1}$ and a fixed $z_M=1$, this can be rewritten as
\begin{equation}
  \label{eq:group_uniring_short_rescaled_2}
  \mathbf{A}^{(k)} z_k \mathbf{c}_0^{(k-1)} = \gamma z_k
  \mathbf{c}_0^{(k)}.
\end{equation}
Substituting $\mathbf{c}^{(k)}=z_{k+1}\mathbf{c}_0^{(k)}$,
Eq.~\eqref{eq:group_uniring_short_rescaled_2} becomes
\begin{equation}
  \label{eq:group_uniring_short_substituted}
  \mathbf{A}^{(k)} \mathbf{c}^{(k-1)} = \gamma \frac{z_k}{z_{k+1}} \mathbf{c}^{(k)}.
\end{equation}
Setting $C^{(k)} = \gamma z_k/z_{k+1}$, it follows that
Eq.~\eqref{eq:group_uniring_short_substituted} yields all possible
solutions of Eq.~\eqref{eq:cond_matrix_form} assuming that
$z_1,\ldots,z_{M-1}$ are free parameters and $z_M=1$.

The scaling factors $z_1,\ldots,z_{M-1}$ change only the magnitude of
the variational vectors, thus their particular choice is not important
for the stability of synchronization. Therefore setting
$\delta\tilde{\mathbf{x}}^{(k)}=z_{k+1}\delta\bar{\mathbf{x}}^{(k)}$
in Eq.~\eqref{eq:group_var_with_linear_combi_cond} yields
\begin{equation}
  \label{eq:group_mse_uniring}
  \delta\dot{\tilde{\mathbf{x}}}^{(k)} = D\mathbf{F}^{(k)}(\mathbf{x}_s^{(k)})\delta\tilde{\mathbf{x}}^{(k)} +
  \gamma \sigma^{(k)} \mathbf{H}^{(k)}\delta\tilde{\mathbf{x}}^{(k-1)}(t-\tau^{(k)}),
\end{equation}
which, in conclusion, qualifies as a master stability equation for
this network topology. Here, $\gamma$ is chosen from the set of
eigenvalues of the block matrix
\begin{equation}
  \label{eq:7}\mathbf{Q} =
  \begin{pmatrix}
    0 & \cdots & \cdots & 0 & \mathbf{A}^{(1)} \\
    \mathbf{A}^{(2)} & 0 & \cdots & \cdots & 0 \\
    0 & \mathbf{A}^{(3)} & 0 & \cdots & 0 \\
    0 & \ddots & \ddots & \ddots & 0 \\
    0 & \cdots & 0 & \mathbf{A}^{(M)} & 0
  \end{pmatrix},
\end{equation}
because Eq.~\eqref{eq:cond_matrix_form_special} is equivalent to the
eigenvalue problem $\mathbf{Q} (\mathbf{c}_0^{(1)},\ldots,\mathbf{c}_0^{(M)}) =
\gamma (\mathbf{c}_0^{(1)},\ldots,\mathbf{c}_0^{(M)})$.

The largest Lyapunov exponent $\Lambda$ calculated from Eq.~(\ref{eq:group_mse_uniring}) 
as a function of the parameter $\gamma \in \mathbb{C}$ is called the 
master stability function (MSF). It determines the stability of group synchronization
if evaluated at the eigenvalues of $\mathbf{Q}$.

\section{Symmetry of the master stability function}
\label{sec:symm-mast-stab}

Note that the master stability equation~\eqref{eq:group_mse_uniring}
($k=1,\ldots,M$) is of dimension $\sum_{k=1}^M d_k$ and thus
independent of the sizes of the individual groups and the particular
coupling topologies $\mathbf{A}^{(k)}$. Because of the structure of 
$\mathbf{Q}$, there always exist $M$ eigenvalues $\gamma_k=\exp(2\pi
ik/M)$ corresponding to dynamics inside the group synchronization
manifold. We will refer to these as longitudinal eigenvalues.

Besides these longitudinal eigenvalues, the spectrum of $\mathbf{Q}$
shows a more general symmetry: For a given eigenvalue $\gamma_j$ of
$\mathbf{Q}$, $\gamma_j \exp(2\pi i k /M)$ is also an eigenvalue of
$\mathbf{Q}$ for any $k=1,\ldots,M$. See
Appendix~\ref{sec:spectr-coupl-matr} for a detailed survey on the
spectrum of the coupling matrix $\mathbf{Q}$.

Looking closely at the master stability
equation~\eqref{eq:group_mse_uniring}, we find another symmetry. The
equation is invariant with respect to the transformation 
$\gamma \rightarrow \exp(-2 \pi i/M) \gamma$:
\begin{eqnarray}
  \label{eq:group_mse_trans}
  \quad \delta\dot{\tilde{\mathbf{x}}}^{(k)} &=&
  D\mathbf{F}^{(k)}(\mathbf{x}_s^{(k)}) \delta\tilde{\mathbf{x}}^{(k)} \\
  &&+
  \gamma \sigma^{(k)}\mathbf{H}^{(k)}e^{\frac{ -2 \pi i}{M}
  }\delta\tilde{\mathbf{x}}^{(k-1)}(t-\tau^{(k)}) \nonumber \\
  \Leftrightarrow e^{\frac{ 2k \pi i}{M} } \delta\dot{\tilde{\mathbf{x}}}^{(k)} &=&
  D\mathbf{F}^{(k)}(\mathbf{x}_s^{(k)})e^{\frac{ 2k \pi i}{M}
  }\delta\tilde{\mathbf{x}}^{(k)} \\
  &&+
  \gamma \sigma^{(k)}\mathbf{H}^{(k)}e^{\frac{ 2(k-1) \pi i}{M}
  }\delta\tilde{\mathbf{x}}^{(k-1)}(t-\tau^{(k)}). \nonumber
\end{eqnarray}
With the basis transformation $\delta\tilde{\mathbf{x}}^{(k)}
\rightarrow \exp(-2k \pi i/M) \delta\tilde{\mathbf{x}}^{(k)}$, which
leaves the Lyapunov spectrum unchanged, the original equation is
regained Consequently, the master stability equation is invariant with
respect to rotations $\gamma \rightarrow \exp(-2 \pi i/M) \gamma$.

Combining both results -- the invariance of the MSF 
and the spectrum of $\mathbf{Q}$ against rotations of $2\pi/M$ -- 
we can conclude that it is sufficient to evaluate the 
MSF in an angular sector given by $\arg(\gamma) \in [0,2\pi/M)$.

In the next Section, we demonstrate this symmetry and calculate the MSF 
for the example of delay-coupled laser networks.

\section{Example: laser networks}\label{sec:laser}

For semiconductor lasers subjected to optical feedback, the
Lang-Kobayashi (LK) model \cite{LAN80b} is a paradigmatic 
model. This model is based on simple rate equations and includes as
variables the carrier inversion $n$ and the complex electric field
$E$, which is reduced to its slowly varying envelope.  The LK model in
its dimensionless form includes the local dynamics
\begin{equation}
  \label{eq:lk_model_vector}
  \mathbf{F}(\mathbf{x}) = \left( \begin{array}{c}
      \frac{1}{T}\left[p-n-\left(1+n\right)\left(x^2+y^2\right)\right] \\
      \frac{n}{2}\left(x-\alpha y\right) \\
      \frac{n}{2}\left(\alpha x + y\right),
    \end{array} \right)
\end{equation}
where $\mathbf{x} =(n,x,y)$ denotes the excess carrier density $n$ and the
complex electric field $E=x+iy$. $T$ denotes the ratio of carrier and
photon lifetimes, $p$ is the normalized pump current in excess of the 
laser threshold, and $\alpha$ is the linewidth enhancement factor. 
The dynamics of a solitary laser --
without any feedback or coupling -- is described by $\dot{\mathbf{x}}
= \mathbf{F}(\mathbf{x}(t))$. Coupling $M$ groups of lasers in a
network of the form Eq.~\eqref{eq:group_diff_eqn}, we consider identical local
dynamics
$\mathbf{F}^{(k)}(\mathbf{x}^{(k)})=\mathbf{F}(\mathbf{x}^{(k)})$ and
focus on all-optical coupling, thus
\begin{equation}\label{eq:H}
  \mathbf{H}^{(k)}=\left(\begin{smallmatrix}
      0 & 0 & 0 \\
      0 & 1 & 0 \\
      0 & 0 & 1
    \end{smallmatrix}\right).
\end{equation}
The cluster synchronization manifold and thus the master stability
equation~\eqref{eq:group_mse_uniring} are $3M$-dimensional.
\begin{figure}
  \includegraphics[width=\linewidth]{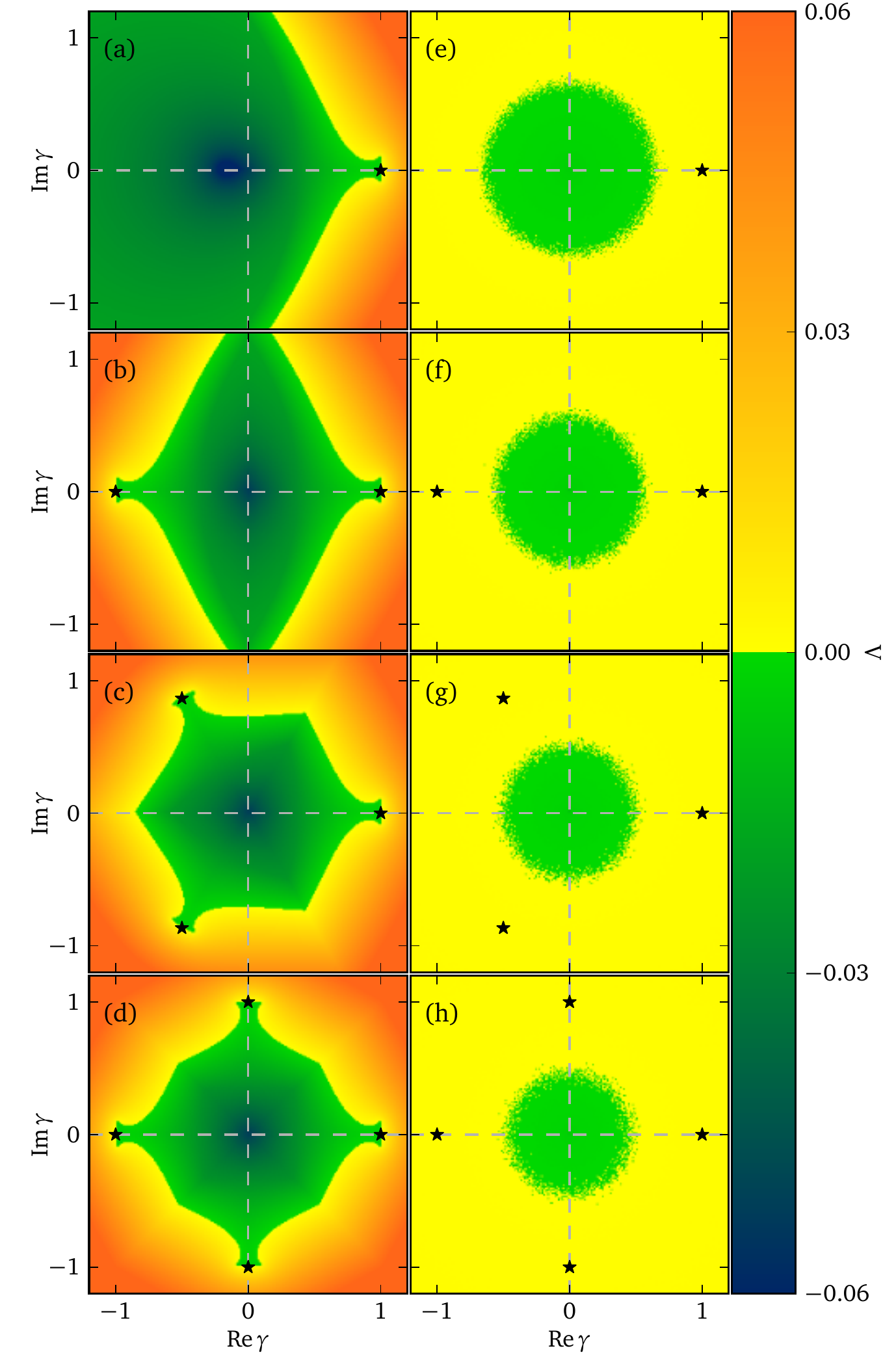}
  \caption{(Color online) Master stability function (MSF) in terms of
    largest Lyapunov exponent $\Lambda(\gamma)$ from
    Eq.~\eqref{eq:group_mse_uniring} for $M=1$, 2, 3, and 4 groups of
    delay-coupled lasers~\eqref{eq:lk_model_vector} in panels (a) and
    (e), (b) and (f), (c) and (g), and (d) and (h),
    respectively. Asterisks mark the position of the longitudinal
    eigenvalues. Left: $\tau^{(k)}\equiv\tau=1$, right:
    $\tau^{(k)}\equiv\tau=1000$. Other parameters:
    $\sigma^{(k)}\equiv\sigma=0.12$, $T=200$, $p=0.1$, $\alpha=4$.}
  \label{fig:msf-groups}
\end{figure}
Figure~\ref{fig:msf-groups} shows the MSF for one, two, three, and
four clusters in panel (a), (b), (c), and (d), respectively. The black
asterisks mark the position of the longitudinal eigenvalues
$\gamma_k=\exp(2\pi ik/M)$, $k=1,\ldots,M$, of $\mathbf{Q}$. Clearly
visible in panels (a)-(d) is the symmetry with respect to discrete
rotations of $2\pi/M$ as discussed in the last Section. In particular,
the Lyapunov exponent $\Lambda(\gamma_k)$ is identical at all
longitudinal eigenvalues $\gamma_k=\exp(2\pi ik/M)$, $k=1,\ldots,M$.
Note that this result is independent of a particular topology.
Choosing any topology that has the structure~\eqref{eq:7}, its
eigenvalue will always show the discrete rotational symmetry discussed
in Sec.~\ref{sec:symm-mast-stab}.

For large delay, as shown in the right part of the Figure, the MSF has
a circular shape for one cluster (panel (e)). This was recently shown
to be a universal feature of networks where the coupling delay is
large compared to the time scale of the local dynamics
\cite{FLU10b}. Due to the discrete rotational symmetry discussed
above, the circular shape cannot change when increasing the number of
clusters, hence the shape of the MSF is independent of the number of
clusters for large coupling delay (see
Fig.~\ref{fig:msf-groups}(f-h)), but we observe that the size of the
disc of stability is shrinking with increasing number of
clusters. This shrinking can be explained as follows: The dimension of
the synchronization manifold Eq.~\eqref{eq:group_sync_manifold} is
proportional to the number of clusters. Since the blocks of the matrix
$\mathbf{Q}$ are arranged in a unidirectional ring, the dynamics
inside the synchronization manifold lives inside such a unidirectional
ring. Hence, the time that a signal takes to travel through this ring
scales linearly with the number of groups $M$. This signal
traveling-time can be seen as an effective time-delay governing the
degree of chaos, i.e., the longitudinal Lyapunov exponent. As was
shown in Refs.~\cite{FLU10b,HEI11}, a larger longitudinal Lyapunov
exponent yields a smaller radius of the stable region.

\begin{figure}
  \includegraphics[width=\linewidth]{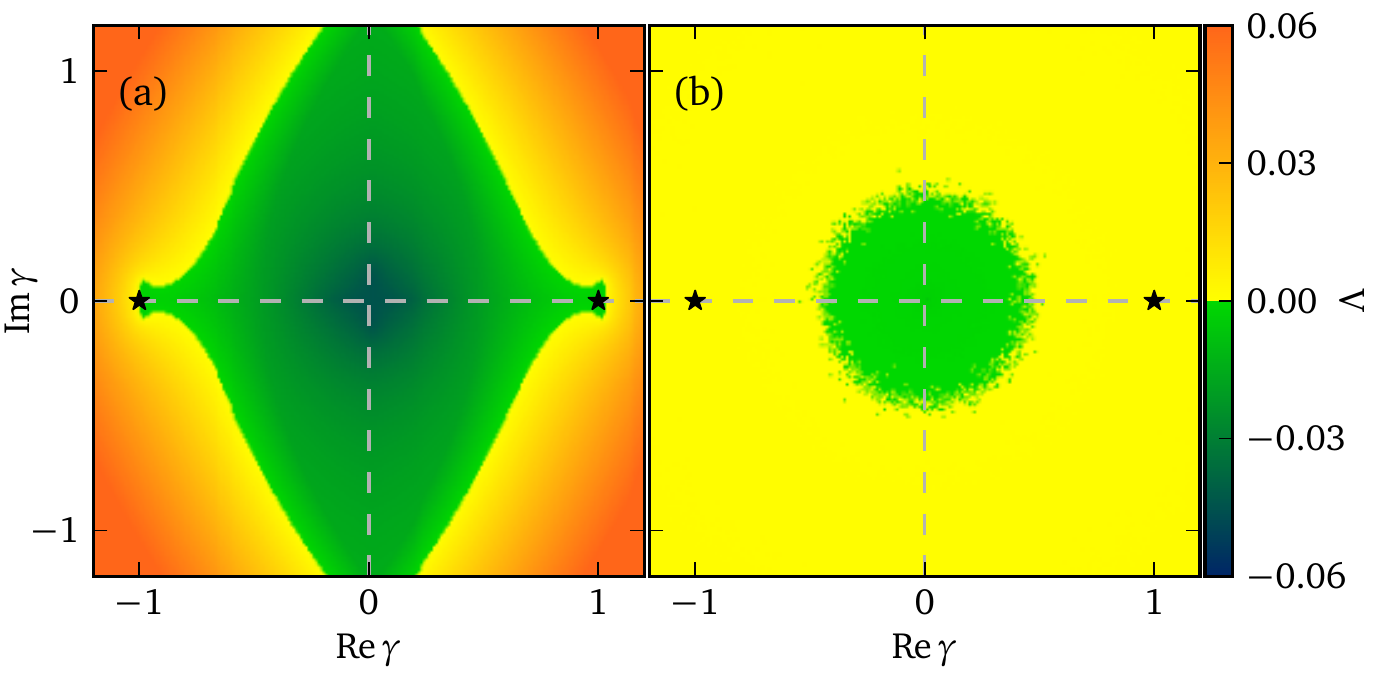}
  \caption{(Color online) Master stability function (MSF) in terms of
    largest Lyapunov exponent $\Lambda(\gamma)$ for two groups of
    delay-coupled lasers~\eqref{eq:lk_model_vector}. The pump current
    is chosen as $p=0.1$ in the first group and $p=0.4$ in the second
    group. (a) $\tau^{(k)}\equiv\tau=1$, (b)
    $\tau^{(k)}\equiv\tau=1000$, other parameters as in
    Fig.~\ref{fig:msf-groups}.}
  \label{fig:lkgroups}
\end{figure}

The above example used identical local dynamics in all of the groups,
which corresponds to the case of cluster synchronization. In order to
illustrate our theory for group synchronization, we now consider two
groups of lasers, where the pump current is $p=0.1$ in the first group
and $p=0.4$ in the second group. Figure~\ref{fig:lkgroups} shows the
resulting master stability function for delay times $\tau=1$ and
$\tau=1000$ in panels (a) and (b), respectively. Compared to
Figs.~\ref{fig:msf-groups}(b,f), only the pump current in one of the
groups is increased. In the case of a small delay time
(Fig.~\ref{fig:lkgroups}(a)) this does not change the master stability
function, because both groups still lock to the same dynamics. In the
case of a large delay time (Fig.~\ref{fig:lkgroups}(b)), the stable
region shrinks compared to Fig.~\ref{fig:msf-groups}(f) due to the
higher pump current in one of the groups leading to a larger
longitudinal Lyapunov exponent $\Lambda(\gamma=\pm 1)$
\cite{FLU10b,HEI11}. Note that the discrete symmetry of the master
stability function is also present for group synchronization.

\section{Beyond multipartite topologies}
\label{sec:beyond-one-block}

So far we have developed a master stability formalism to determine the
stability of group and cluster synchronization.  In order to utilize
the master stability framework, one major restriction has been made: Each
group must receive input from one and only one other group, i.e., the
network topology has to be multipartite. More complex topologies beyond
multipartite structures like, for instance, lattices
\cite{KES07,KES08} could not be dealt with. In the following we will
derive the master stability equation for group synchronization with
multiple coupling matrices. Thereby some of the former stringent restrictions 
can be dropped.

The network dynamics for group synchronization with one coupling
matrix has been written in the form of Eq.~\eqref{eq:group_diff_eqn}, and the
synchronization manifold and the master stability equation were given
by Eqs.~\eqref{eq:group_sync_manifold}
and~\eqref{eq:group_mse_uniring}, respectively. Generalizing this to
two coupling matrices yields for the network dynamics of $M$ groups:
\begin{align}
  \label{eq:79}
  \dot{\mathbf{x}}_i^{(k)} = \mathbf{F}^{(k)}[\mathbf{x}_i^{(k)}(t)]
  &+ \sigma^{(k)}_A \sum_{j=1}^{N_n} A^{(k)}_{ij}
  \mathbf{H}^{(k)}\mathbf{x}_j^{(k-1)}(t-\tau^{(k)})\nonumber \\ &+
  \sigma^{(k)}_B \sum_{j=1}^{N_n} B^{(k)}_{ij}
  \mathbf{H}^{(k)}\mathbf{x}_j^{(n_k)}(t-\tau^{(k)}),
\end{align}
where the matrix $\mathbf{A}^{(k)}$ describes the coupling from the
$(k-1)$-th to the $k$-th group as before and $\mathbf{B}^{(k)}$
describes the coupling from the $n_k$-th to the $k$-th group. That is,
the $k$-th group now receives input from two groups, $k-1$ and $n_k$. The
row sums of all $\mathbf{A}^{(k)}$ and $\mathbf{B}^{(k)}$ must be
unity. Any constant non-zero row sum can be rescaled by means of the coupling
strengths.

For the sake of simplicity and readability, we use identical coupling
schemes and identical time delays for both coupling terms. In general,
our framework works for different time delays and coupling schemes.
The sum of $\sigma^{(k)}_A$ and $\sigma^{(k)}_B$ must yield the
overall coupling strength $\sigma^{(k)}$ used before in order to
arrive at the same dynamical regime:
$\sigma^{(k)}_A+\sigma^{(k)}_B=\sigma^{(k)}$.
Figure~\ref{fig:schema}(a) shows schematically the coupling parameters
and matrices that are present in Eq.~\eqref{eq:79}. In the case of two
groups shown here, $\mathbf{B}^{(1)}$ and $\mathbf{B}^{(2)}$ represent
the coupling within the groups, depicted by solid arrows.

From the above, the synchronization manifold is obtained as
\begin{align}
  \label{eq:80}
  \dot{\mathbf{x}}_s^{(k)} = \mathbf{F}^{(k)}[\mathbf{x}_s^{(k)}(t)] &+
  \sigma^{(k)}_A \mathbf{H}^{(k)}\mathbf{x}_s^{(k-1)}(t-\tau^{(k)}) \nonumber\\
  &+
  \sigma^{(k)}_B \mathbf{H}^{(k)}\mathbf{x}_s^{(n_k)}(t-\tau^{(k)})
\end{align}
for $k=1,\ldots,M$. See Fig.~\ref{fig:schema}(b) for a schematic
diagram of the synchronization manifold for the example of two groups.
The coupling inside a group translates into a self-feedback loop,
depicted by solid arrows. Let $\mathbf{Q}_A$ be the matrix containing
the blocks $\mathbf{A}^{(kn)}$ at positions $(k,k-1)$ and
$\mathbf{Q}_B$ the matrix containing the blocks $\mathbf{B}^{(k)}$ at
positions $(k,n_k)$. If $\mathbf{Q}_A$ and $\mathbf{Q}_b$ commute,
i.e., $[\mathbf{Q}_A,\mathbf{Q}_B]=0$, it is possible to obtain a
master stability equation
\begin{align}
  \label{eq:82}
  \delta\dot{\bar{\mathbf{x}}}^{(k)} =& \ 
  \mathbf{DF}^{(k)}(\mathbf{x}_s^{(k)})\delta\bar{\mathbf{x}}^{(k)}(t)
  \nonumber \\
  &\quad + \sigma^{(k)}_A
  \gamma^{(1)}
  \mathbf{H}^{(k)}\delta\bar{\mathbf{x}}^{(k-1)}(t-\tau^{(k)})
  \nonumber \\ &\quad + \sigma^{(k)}_B
  \gamma^{(2)} \mathbf{H}^{(k)}\delta\bar{\mathbf{x}}^{(n_k)}(t-\tau^{(k)}),
\end{align}
for $k=1,\ldots,M$, where $\gamma^{(1)}$ and $\gamma^{(2)}$ are chosen
from the eigenvalue spectrum of the matrices matrices $\mathbf{Q}_A$
and $\mathbf{Q}_B$, respectively. These eigenvalues have to be
evaluated in pairs corresponding to one eigenvector. Since
$\mathbf{Q}_A$ and $\mathbf{Q}_B$ commute they always have a set of
identical eigenvectors.

\subsection{Example: two groups}

\begin{figure}[t]
  \centering
  \includegraphics[width=\linewidth]{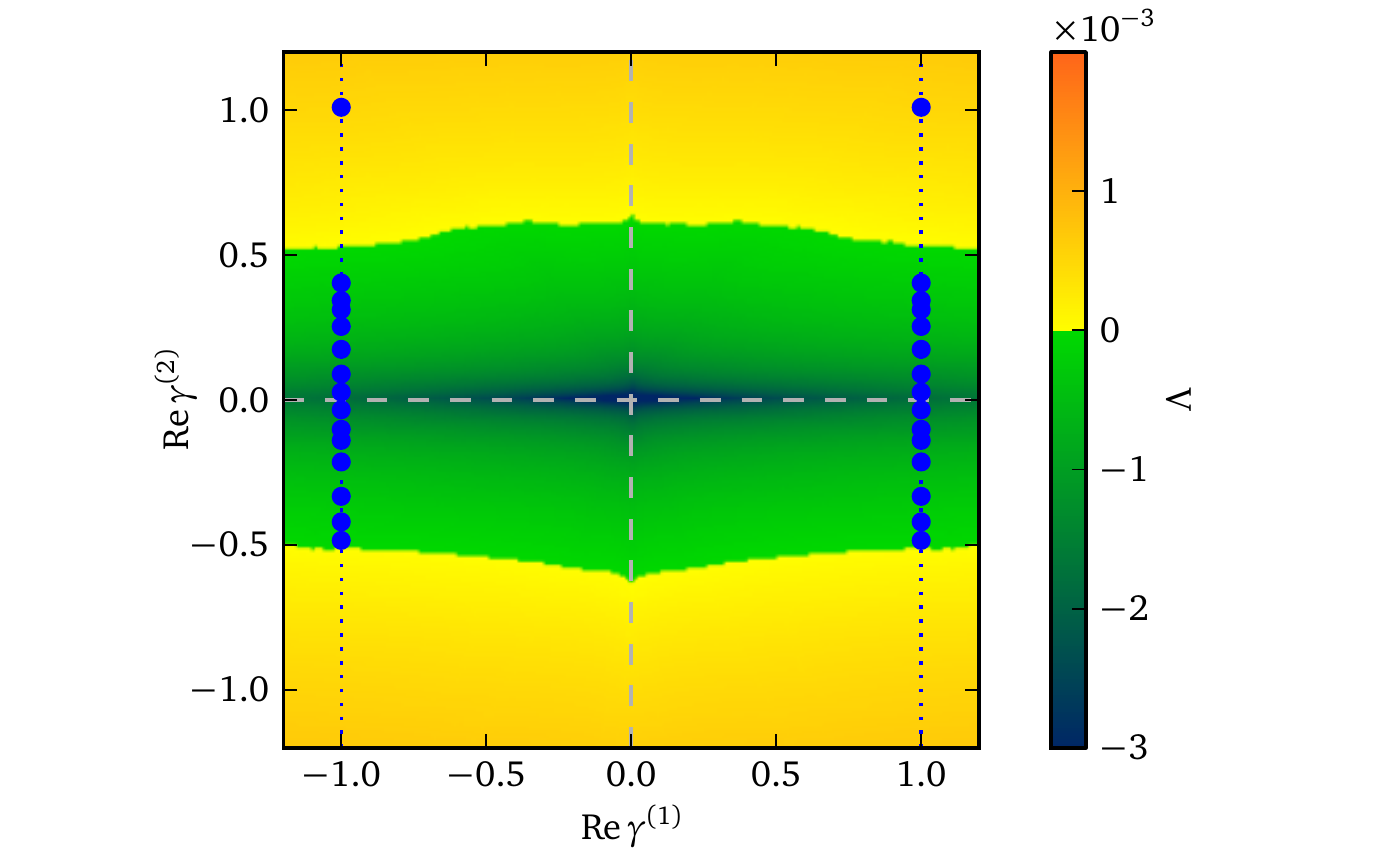}
  \caption{(Color online) Master stability function for two commuting
    matrices with the structures $\mathbf{Q}_A$ as in
    Eq.~\eqref{eq:34} and $\mathbf{Q}_B$ as in Eq.~\eqref{eq:36} with
    coupling strengths $\sigma^{(1)}_A=\sigma^{(2)}_A=0.05\sigma$ and
    $\sigma^{(1)}_B=\sigma^{(2)}_B=0.95\sigma$ with $\sigma=0.12$. The
    pairs $(\nRe\gamma^{(1)},\nRe\gamma^{(2)})$ plotted as black (blue) dots
    correspond to eigenvalues of the hierarchical network with
    matrices~\eqref{eq:124} and~\eqref{eq:125} using a link
    probability $p=0.5$ in the Erd\H{o}s-R\'{e}nyi
    graph~\eqref{eq:125}. Other parameters: $T=200$, $p=0.1$,
    $\alpha=4$, $\tau=1000$.}
\label{fig:zwei_2Q_hierarch_r005}
\end{figure}

Let us first consider the simplest example, namely only two
groups. Above, we have shown results for synchronization in two groups
using a single coupling matrix of the form
\begin{equation}
  \label{eq:34}
  \mathbf{Q}_A =
  \begin{pmatrix}
    0 & \mathbf{A}^{(1)} \\
    \mathbf{A}^{(2)} & 0
  \end{pmatrix},
\end{equation}
where the matrices $\mathbf{A}^{(1)}$ and $\mathbf{A}^{(2)}$ describe
the coupling from the second to the first group and vice versa,
respectively, see also Ref.~\cite{SOR07}. We will now elaborate what
happens when we introduce a second coupling matrix
\begin{equation}
  \label{eq:36}
    \mathbf{Q}_B =
  \begin{pmatrix}
    \mathbf{B}^{(1)} & 0\\
    0 &\mathbf{B}^{(2)}
  \end{pmatrix},
\end{equation}
i.e., $n_1=1$ and $n_2=2$. Figure~\ref{fig:zwei_2Q_hierarch_r005}
shows the master stability function for the structure given by these
matrices $\mathbf{Q}_A$ and $\mathbf{Q}_B$ and for laser parameters in
the regime of low-frequency fluctuations with $T=200$, $p=0.1$,
$\alpha=4$. The coupling strengths are chosen as
$\sigma^{(1)}_A=\sigma^{(2)}_A=0.05\sigma$ and
$\sigma^{(1)}_B=\sigma^{(2)}_B=0.95\sigma$ with $\sigma=0.12$. This
resembles strong coupling in the clusters, but weak coupling between
clusters. We use only one time delay $\tau=1000$ for simplicity. For
the same reason, we investigate cluster synchronization, i.e., the
local dynamics $\mathbf{F}$ and the coupling scheme $\mathbf{H}$ are
identical for both groups in this example. We consider matrices with
real eigenspectrum only and set
$\nIm\gamma^{(1)}=\nIm\gamma^{(2)}=0$. The eigenvalue pairs depicted
by the black (blue) dots correspond to a particular network topology that will
be discussed below.

Using the forms~\eqref{eq:34} and~\eqref{eq:36}, the commutation
relation $[\mathbf{Q}_A,\mathbf{Q}_B]=0$ is equivalent to
\begin{equation}
  \label{eq:106}
  \left\{ \begin{array}{l}
      \mathbf{A}^{(1)}\mathbf{B}^{(2)} = \mathbf{B}^{(1)}\mathbf{A}^{(1)}\\
      \mathbf{A}^{(2)}\mathbf{B}^{(1)} = \mathbf{B}^{(2)}\mathbf{A}^{(2)}.
    \end{array}
  \right.
\end{equation}

These conditions are fulfilled for certain classes of matrices
only. 
We will give an example of hierarchical coupling that
yields matrices which fulfill these conditions.

\subsection{Towards hierarchical networks}
\label{sec:towards-hier-netw}

A hierarchical network usually consists of topological clusters that
are densely coupled inside, while links to other such topological
clusters are sparse. The hierarchy is then built by larger topological
clusters that contain the smaller ones \cite{ZHO06c,ZHO06d}. This
procedure can be continued over many levels of hierarchy. It is
important to distinguish these topological clusters from the dynamical
cluster states that are investigated in this paper.

\begin{figure}[t]
  \centering
  \includegraphics[angle=270,width=0.6\linewidth]{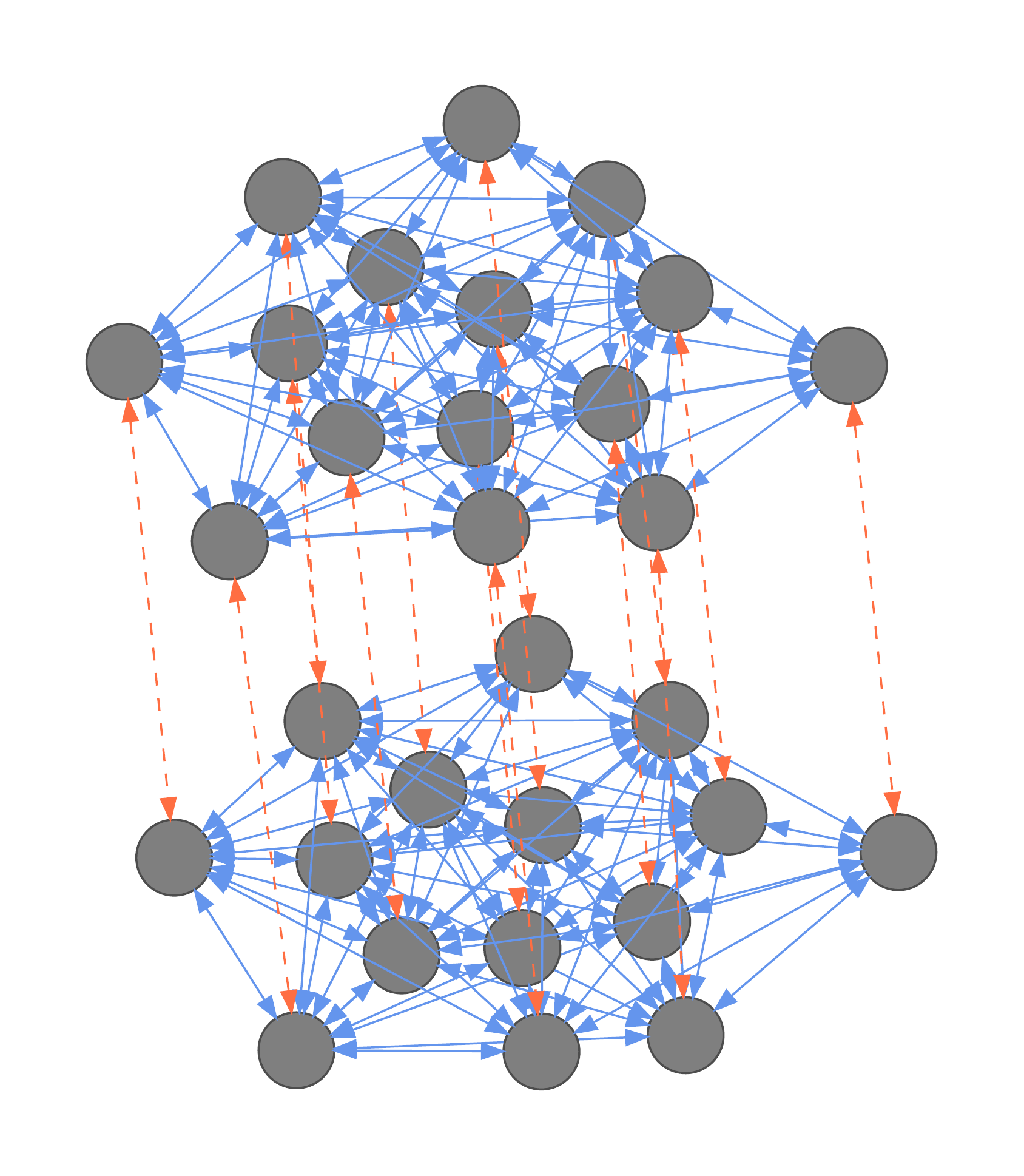}
  \caption{(Color online) Schematic view of a simple hierarchical
    network structure according to Eqs.~\eqref{eq:124}
    and~\eqref{eq:125} with $N=30$ nodes. The two topological clusters
    are separated for illustration. Solid (blue) and dashed (red)
    arrows correspond to links inside ($\mathbf{Q}_B$) and between
    ($\mathbf{Q}_A$) the clusters, respectively.}
  \label{fig:hierarch_schema}
\end{figure}

The simplest hierarchical structure consists of just two topological
clusters. Figure~\ref{fig:hierarch_schema} illustrates this in a
schematic sketch of a graph of $N=30$ nodes with two topological
clusters, $N_1=N_2=N/2$. Solid (blue) arrows correspond to links
inside one cluster while dashed (red) arrows denote links between both
clusters.
In this Section we will show that each cluster can exhibit isochronous
synchronization under certain conditions. In this sense,
the notions of topological cluster and of dynamical cluster coincide at this
point.

The graph in Fig.~\ref{fig:hierarch_schema} is modeled by the coupling
matrices
\begin{equation}
  \label{eq:124}
  \mathbf{Q}_A =
  \begin{pmatrix}
    0 & \mathbf{1}_{N/2} \\
    \mathbf{1}_{N/2} & 0
  \end{pmatrix}
\end{equation}
and
\begin{equation}
  \label{eq:125}
  \mathbf{Q}_B =
  \begin{pmatrix}
    \mathbf{B} & 0\\
    0 &\mathbf{B}
  \end{pmatrix},
\end{equation}
where $\mathbf{1}_{N/2}$ is the identity matrix and $\mathbf{B}$ is an
undirected $N/2 \times N/2$ Erd\H{o}s-R\'enyi random graph with a
certain link probability $p$. The undirectedness is necessary to
obtain a real-valued eigenvalue spectrum. Then it is sufficient to
calculate the master stability function in the
$(\nRe\gamma^{(1)},\nRe\gamma^{(2)})$ plane as done in
Fig.~\ref{fig:zwei_2Q_hierarch_r005}.

In order to comply with the link density of a hierarchical network, we
choose the coupling strength for the two matrices $\mathbf{Q}_A$ and
$\mathbf{Q}_B$ to be different as used for the calculation of the
master stability function in Fig.~\ref{fig:zwei_2Q_hierarch_r005}. The
coupling strengths $\sigma^{(1)}_A$ and $\sigma^{(2)}_A$ are chosen as
$\sigma^{(1)}_A=\sigma^{(2)}_A=0.05\sigma$, where $\sigma=0.12$ is the
overall coupling strength corresponding to the regime of low-frequency
fluctuations of the laser dynamics. The coupling strengths corresponding to $\mathbf{Q}_B$
are chosen as $\sigma^{(1)}_B=\sigma^{(2)}_B=0.95\sigma$. For a high
link probability $p$ in the random matrix $\mathbf{B}$, the matrix
$\mathbf{Q}_B$ contains comparatively more links than $\mathbf{Q}_A$,
which has only one link per row. Given that both matrices are
renormalized to unity row sum, it is therefore a reasonable choice
that $\sigma^{(1)}_B$ and $\sigma^{(2)}_B$ are significantly larger.

The black (blue) dots in Fig.~\ref{fig:zwei_2Q_hierarch_r005} show the
eigenvalue pairs $(\gamma^{(1)},\gamma^{(2)})$ of the hierarchical
example given by Eqs.~\eqref{eq:124} and~\eqref{eq:125} using a link
probability of $p=0.5$ in the random matrix $\mathbf{B}$ for
$N=30$. It can be seen that this network shows stable synchronization
in this 2-cluster state. That is, each topological cluster exhibits
synchronization internally. All eigenvalue pairs transversal to the
synchronization manifold are inside the stable region, while the
longitudinal eigenvalue pairs $(1,1)$ and $(-1,1)$ do not affect the
stability of synchronization. For any choice of $\mathbf{Q}_B$ the
eigenvalues will always be lined up on the dotted vertical lines,
which are determined by the matrix $\mathbf{Q}_A$ being constructed from
identity-matrix blocks.

The link probability $p=0.5$ is just above the threshold of stable
synchronization. Using lower values, some eigenvalues will cross the
boundary of the stable region of the master stability function,
leading to desynchronization.

Since the eigenvalues are always aligned along the lines
$\nRe{\gamma^{(1)}}=\pm 1$ in this example, the stability for other
choices of the coupling strength can easily be obtained by evaluating
the master stability function at a fixed value of
$\nRe{\gamma^{(1)}}=1$ as a function of $\nRe{\gamma^{(2)}}$ and
$\sigma_A$ or $\sigma_B$. The other value $\nRe{\gamma^{(1)}}=-1$
yields identical results and therefore need not be considered, which
is a result of the symmetry discussed in
Sec.~\ref{sec:symm-mast-stab} and observable also in
Fig.~\ref{fig:zwei_2Q_hierarch_r005}.

\begin{figure}[t]
  \centering
  \includegraphics[width=\linewidth]{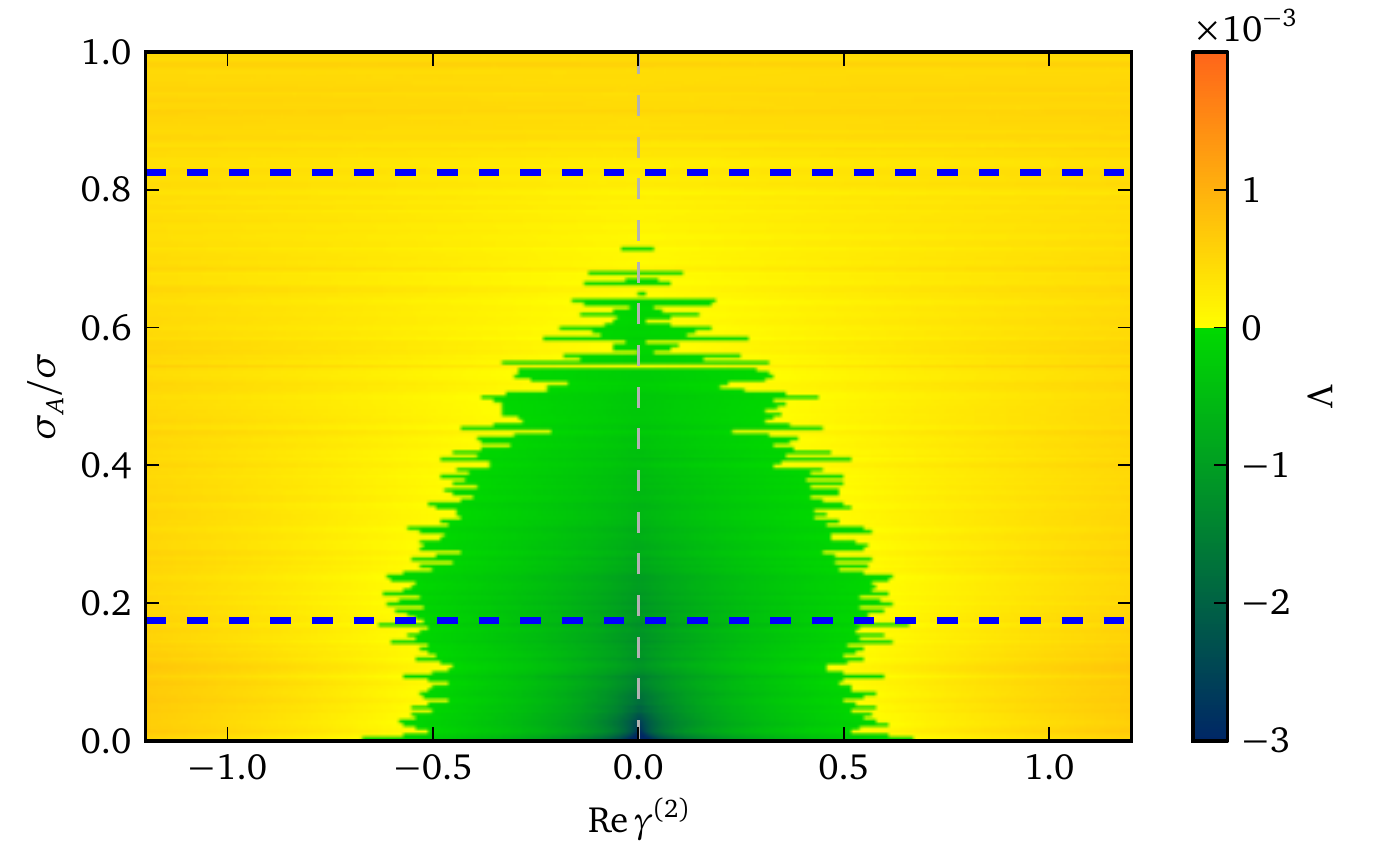}
  \caption{(Color online) Master stability function for two commuting
    matrices with the structures $\mathbf{Q}_1$ as in
    Eq.~\eqref{eq:34} and $\mathbf{Q}_2$ as in Eq.~\eqref{eq:36} in
    the $(\nRe{\gamma^{(2)}},\sigma_A/\sigma)$ plane.
    $\nRe{\gamma^{(1)}}=1$, $\nIm{\gamma^{(1)}}=\nIm{\gamma^{(2)}}=0$
    $\sigma_B = \sigma-\sigma_A$, $\sigma=0.12$. The dashed black (blue) lines
    form the boundary of the parameter range where no 2-cluster
    exists. Parameters: $T=200$, $p=0.1$, $\alpha=4$, $\tau=1000$.}
\label{fig:zwei_2Q_Revsrowsum}
\end{figure}
Figure~\ref{fig:zwei_2Q_Revsrowsum} shows the master stability
function in the $(\nRe{\gamma^{(2)}},\sigma_A/\sigma)$ plane, where
$\sigma_A \equiv\sigma^{(1)}_A=\sigma^{(2)}_A$. The other coupling strength
is set using the relation $\sigma_B\equiv\sigma^{(1)}_B=\sigma^{(2)}_B=
\sigma-\sigma_A$, where the overall coupling strength is chosen as
$\sigma=0.12$ corresponding to the laser regime of low-frequency
fluctuations. This relation ensures that the overall coupling strength
leads to an operation in this regime.

The dashed black (blue) lines enclose the region where no 2-cluster state can
exist. This can be seen by considering the two-node network motif
described by
\begin{equation}
  \label{eq:6}
  \mathbf{x}_s^{(k)}=\mathbf{F}[\mathbf{x}_s^{(k)}] + \sum_{j=1}^2
  \hat{G}_{kj} \mathbf{H} \mathbf{x}_s^{(j)}(t-\tau),
\end{equation}
$k=1,2$, where the coupling matrix
\begin{equation}
  \label{eq:122}
  \hat{\mathbf{G}} = \begin{pmatrix}
    \sigma^{(1)}_B & \sigma^{(1)}_A \\
    \sigma^{(2)}_A & \sigma^{(2)}_B 
  \end{pmatrix},
\end{equation}
describes the behavior on the 2-cluster synchronization manifold. For
coupling strengths between the black (blue) lines, this motif shows stable
synchronization for the chosen laser parameters and thus the dynamics
in both clusters will be identical. The stability of the 2-cluster
state is therefore only meaningful below the lower and above the upper
dashed black (blue) line.

The boundaries of stability for the 2-cluster state are nearly
independent of $\sigma_1/\sigma$ in the lower range of
$\sigma_1/\sigma<0.175$, which corresponds to a high coupling strength
inside the clusters, but a low coupling strength between clusters. The
upper range of $\sigma_1/\sigma>0.825$, which corresponds to low
coupling strength inside the clusters, but high coupling strength
between them, also allows for the existence of the 2-cluster state,
but this state cannot be stable for any topology.  In conclusion, the
coupling strength must be comparatively large inside the clusters to
allow for a stable 2-cluster state.

\section{Example: neural networks}
\label{sec:exampl-neur-netw}
Synchronization in the brain can be related to cognitive capacities
\cite{SIN99b} as well as to pathological conditions, e.g., epilepsy
\cite{UHL06}. Therefore, there has been tremendous interest in the
study of synchronization in neural networks \cite{COO09, LU04,
  TIM06,UHL09}. The master stability approach has been applied to the
study of synchronization patterns independently of a specific network topology
\cite{DHA04, JIR08, LEH11}. The brain is organized in different brain
areas leading to different delay times between neurons of different
areas and neurons within the same area. Furthermore, different types of 
neurons exist, corresponding to different local dynamics. Therefore
we propose that the master stability function for group synchronization
introduced here will be especially useful for investigating complex 
neural synchronization phenomena.

Here we apply our method to a neural network where the nodes are
modeled as FitzHugh-Nagumo (FHN) systems. As in the last Section we
consider a network of two groups coupled via two coupling matrices
$\mathbf{Q}_A$ (intergroup coupling) and $\mathbf{Q}_B$ (intragroup
coupling). We use a diffusive-like coupling. As discussed above this
can be transformed to the coupling used in the previous sections by
transforming the local dynamics of the $i$-th node in the $k$-th 
cluster as follows:
\begin{eqnarray}
  \label{eq:fhn}
  \mathbf{F}(\mathbf{x}_i^{(k)}) &=& \left( \begin{array}{c}
      \frac{1}{\epsilon}  (u_i^{(k)} - \frac{1}{3}{u_i^{(k)}}^3 - v_i ^{(k)})    \\
      u_i^{(k)}+a 
    \end{array} \right) \nonumber \\
  & & \quad + (\sigma_A^{(k)}+   \sigma_B^{(k)} )
  \mathbf{H} \mathbf{x}_i^{(k)},
\end{eqnarray}
with $\mathbf{x}_i^{(k)}=(u_i^{(k)},v_i^{(k)})$ and $k=1,2$. Here $u$ and
$v$ denote the activator and inhibitor variables, respectively. The parameter $a$
determines the threshold of excitability. A single FHN oscillator is
excitable for $a>1$ and exhibits self-sustained periodic firing beyond
the Hopf bifurcation at $a=1$. We will focus on the excitable
regime with $a=1.3$. The time-scale parameter $\epsilon$ is chosen as
$\epsilon = 0.01$. The synchronized dynamics and the master stability
equation are then given by Eq.~\eqref{eq:80} and Eq.~\eqref{eq:82},
respectively. We assume the coupling scheme $\mathbf{H}^{(1)}=\mathbf{H}^{(2)}\equiv \mathbf{H}
=\left(\begin{smallmatrix}1/\epsilon & 0\\0 &
    0\end{smallmatrix}\right)$.

\begin{figure}[t]
  \centering
  \includegraphics[width=\linewidth]{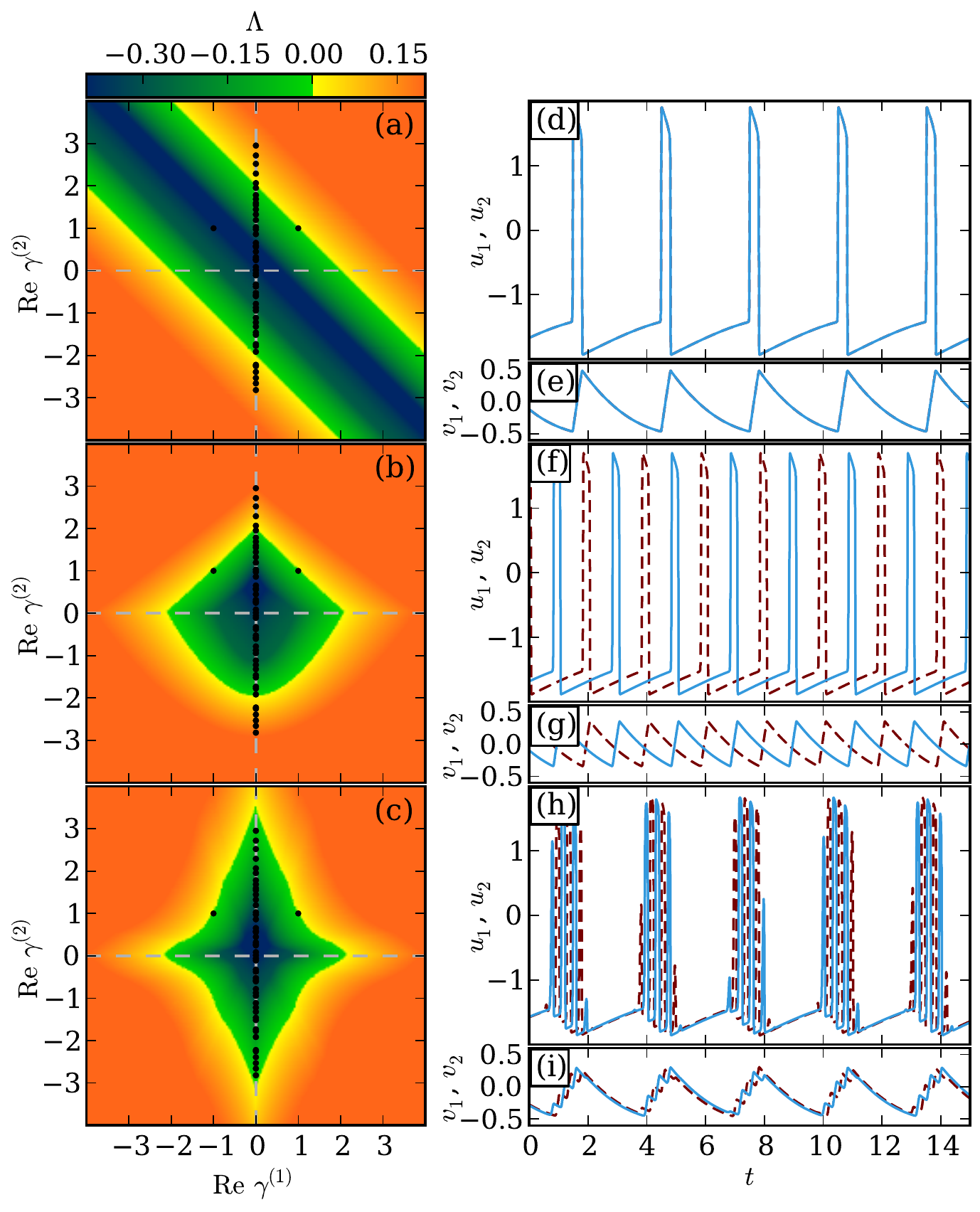}
  \caption{(Color online) (a)-(c): Master stability function for
    networks of FitzHugh-Nagumo oscillators governed by
    Eq.~\eqref{eq:fhn} in the
    $(\nRe{\gamma^{(1)}},\nRe{\gamma^{(2)}})$ plane for
    $\nIm{\gamma^{(1)}}=\nIm{\gamma^{(2)}}=0$ and different delay
    times. The black dots denote the location of the eigenvalue pairs
    for the example topology~\eqref{eq:QB}. (d)-(i): Time series of
    the dynamics in the first (dark dashed red) and second (light
    solid blue) group. Parameters: (a),(d),(e): in-phase
    synchronization ($ \tau_B^{(k)}=3$), (b),(f),(g): anti-phase
    synchronization ($ \tau_B^{(k)}=2$), (c),(h),(i): synchronized
    bursting ($ \tau_B^{(k)}=3.2$). Other Parameters: $\sigma_A^{(k)}=
    \sigma_B^{(k)} =0.5, \tau_A^{(k)}=3$, $\epsilon=0.01$, $a=1.3$
    (groups $k=1,2$). }
\label{fig:fhn}
\end{figure}

The cluster synchronized dynamics is equivalent to a system of two
coupled nodes with self-feedback. In Ref.~\cite{SCH08} it was shown
that depending on the delay times, the coupling strength, and the
strength of the self-feedback different dynamical scenarios, i.e.,
in-phase synchronization, anti-phase synchronization, or bursting can
arise. Figure~\ref{fig:fhn} shows the master stability function in
panels (a)-(c) for in-phase synchronization, anti-phase
synchronization and for synchronization in two bursting groups,
respectively. The right hand panels of Fig.~\ref{fig:fhn} depict the
corresponding time series: In panel (d), (f), and (h) for the
activator variables and in panel (e), (g), and (i) for the inhibitor
for in-phase, anti-phase, and bursting dynamics, respectively. Because
the different dynamical scenarios yield distinctively different stable
regions, topologies might arise which show stable synchronization for
one of the patterns but not for the others. However, for all scenarios
the stable region contains the unity square, i.e., $(\gamma_1,
\gamma_2) \in [-1,1]\times [-1,1]$. With Gershgorin's circle theorem
\cite{EAR03} it can easily be shown that the eigenvalues of
symmetrical matrices with positive entries and unity row sum are
always contained in the interval $[-1,1]$. Thus, if $\mathbf{Q}_A$ and
$\mathbf{Q}_B$ have only positive entries, i.e., if the coupling is
excitatory, synchronization is stable for the dynamics and parameters
shown here. As a consequence, only the introduction of inhibitory
links can lead to desynchronization. In Ref.~\cite{LEH11} it has been
shown that for $\sigma_A^{(k)}= \sigma_B^{(k)} =\sigma$ and
$\tau_A^{(k)}=\tau_B^{(k)}=\tau$ this is the case for all $\sigma$ and
$\tau$ for which the synchronized dynamics is periodic. A detailed
study of these phenomena for the eight-dimensional parameter space of
$\sigma_A^{(k)} ,\sigma_B^{(k)}, \tau_A^{(k)},\tau_B^{(k)}$ ($ k=1,2$)
is beyond the scope of this paper. Note that the symmetry discussed in
Sec.~\ref{sec:symm-mast-stab} does not show up in Fig.~\ref{fig:fhn}(a),
because both clusters synchronize to
$\mathbf{x}_s^{(1)}=\mathbf{x}_s^{(2)}$ and the invariance of
Eq.~\eqref{eq:group_mse_trans} does not hold in this case of in-phase
synchronized spiking.

As an example of a network with inhibitory links which will exhibit stable
synchronization only in one of the patterns discussed above, but not in
the other ones, we choose $\mathbf{Q}_A$ and $\mathbf{Q}_B$ as 
\begin{equation}
  \label{eq:QB}
  \mathbf{Q}_A =
  \begin{pmatrix}
    0 & \mathbf{A}\\
    \mathbf{A} & 0
  \end{pmatrix},\ 
  \mathbf{Q}_B =
  \begin{pmatrix}
    \mathbf{B} & 0\\
    0 &\mathbf{B}
  \end{pmatrix},
\end{equation}
where $\mathbf{A}={a_{ij}}$ with $a_{ij}=1$ $\forall i,j=1,\ldots,N$ is
an all-to-all coupling matrix with self-coupling, and $\mathbf{B}$ is an
undirected random matrix with both excitatory (positive entries) and
inhibitory links (negative entries). The matrix $\mathbf{B}$ describes a 
fixed node degree with 12 excitatory and 9 inhibitory links for each node. The number of
nodes is chosen as $N=100$. The black dots in Fig.~\ref{fig:fhn}
denote the corresponding eigenvalue pairs. In panels (a) and (b) some
eigenvalues are located outside the stable region, while in panel (c)
they are all inside, which means that the zero-lag and
anti-phase synchronized solutions will be unstable in such a network,
while synchronization in the bursting state will be stable.

\section{Conclusion}\label{sec:conclusion}
Based on a master stability approach, we have studied patterns of cluster
and group synchronization in delay-coupled networks and determined their 
stability. We have shown that the master stability function applied to cluster and group
synchronization exhibits a discrete $M$-fold rotational symmetry 
for $M$ dynamical clusters. This reduces the numerical effort, such 
that for a larger number of clusters the master stability function must be
evaluated only on a smaller angular sector in the
complex plane. Within our approach we can treat a wide range of multipartite network topologies. Using multiple commuting coupling matrices,
we have generalized our stability analysis beyond multipartite topologies, for 
instance towards hierarchical network structures.
As concrete examples we have focused on delay-coupled lasers and neural networks.
The interplay of complex topologies, multiple delay times, and possibly different local
dynamics and different coupling functions extends the
scope of the master stability framework and is a step towards
understanding complex patterns of synchronization in real-world
networks.

\begin{acknowledgments}
 This work was supported by DFG in the framework of SFB 910.
\end{acknowledgments}

\appendix
\section{Spectrum of the coupling matrix}
\label{sec:spectr-coupl-matr}

We investigate the eigenvalue spectrum of the coupling matrix
\begin{equation}
  \label{eq:8}\mathbf{Q} =
  \begin{pmatrix}
    0 & \cdots & \cdots & 0 & \mathbf{A}^{(1)} \\
    \mathbf{A}^{(2)} & 0 & \cdots & \cdots & 0 \\
    0 & \mathbf{A}^{(3)} & 0 & \cdots & 0 \\
    0 & \ddots & \ddots & \ddots & 0 \\
    0 & \cdots & 0 & \mathbf{A}^{(M)} & 0
  \end{pmatrix}.
\end{equation}
$\mathbf{Q}$ has at least $n_0$ zero eigenvalues, where
\begin{equation}
  \label{eq:2}
  n_0 = \sum_k | N_k-N_{k-1} | 
\end{equation}
arises solely due to the block structure of $\mathbf{Q}$: if all
$\mathbf{A}^{(k)}$ have maximum rank $\min(N_k,N_{k-1})$, there are
exactly those $n_0$ zeros. In general, the exact number of zeros is
given by $\sum_k \left(N_k - \rank \mathbf{A}^{(k)}\right)$, which may
be larger than $n_0$ due to the particular structure of the
$\mathbf{A}^{(k)}$.

Consider the matrix $\mathbf{Q}^M$, which is of block
diagonal structure
\begin{widetext}
\begin{equation}
  \label{eq:9}\mathbf{Q}^M =
  \begin{pmatrix}
    \mathbf{A}^{(1)}\mathbf{A}^{(M)}\mathbf{A}^{(M-1)}\cdots\mathbf{A}^{(2)} & 0 & \cdots & 0 \\
    0 & \mathbf{A}^{(2)}\mathbf{A}^{(1)}\mathbf{A}^{(M)}\mathbf{A}^{(M-1)}\cdots\mathbf{A}^{(3)} & \ddots & 0 \\
    0 & \ddots & \ddots & 0 \\
    0 & \cdots & 0 & \mathbf{A}^{(M)}\mathbf{A}^{(M-1)}\cdots\mathbf{A}^{(1)}
  \end{pmatrix}.
\end{equation}
\end{widetext}
Note that each block on the diagonal is a product of all
$\mathbf{A}^{(k)}$, only the order differs.

Assume that the groups are arranged such that $N_1 \leq N_j$
($j=2,\ldots,M$), which can always be achieved by an index permutation, and
that each $\mathbf{A}^{(k)}$ has maximum rank
$\min(N_k,N_{k-1})$\footnote{The latter assumption simplifies the
  argument regarding the zero eigenvalues, but the final result
  is valid for arbitrary ranks of the block matrices}. Then of the
blocks in $\mathbf{Q}^M$, $(\mathbf{Q}^M)_{11}$ has lowest rank, since
it is an $N_1\times N_1$ matrix. The non-zero eigenvalues of a matrix
product are invariant against exchange of the factors, their number
(including degeneracy) equals the rank of the product with lowest
rank, i.e., $(\mathbf{Q}^M)_{11}$ in our case. As a consequence, the
non-zero eigenvalues of $\mathbf{Q}^M$ are given by the non-zero
eigenvalues $\{\lambda_1,\ldots,\lambda_{N_1}\}$ of $(\mathbf{Q}^M)_{11}$. 
As there are $M$ blocks yielding exactly these eigenvalues, each of them is $M$-fold degenerate. In
particular, since the row sum of $\mathbf{Q}^M$ is unity, there is an
$M$-fold unity eigenvalue.

The non-zero eigenvalues of $\mathbf{Q}$ are then given by the $M$-th
roots of the non-zero eigenvalues of $\mathbf{Q}^M$, and the whole
spectrum $\Gamma=\{\gamma_j\}_{j=1,\ldots,\sum N_k}$ of $\mathbf{Q}$
reads
\begin{eqnarray}
  \label{eq:3}
  \Gamma &=& \{\underbrace{0,\ldots,0}_{n_0}\} \cup \bigcup_{k=1}^M \{
  \sqrt[M]{|\lambda_1|} e^{[\arg(\lambda_1) + 2\pi k]
      i/M},\ldots\\
  && \qquad\qquad\qquad\qquad\ldots,\sqrt[M]{|\lambda_M|} e^{[\arg(\lambda_M) +
      2\pi k] i/M} \}. \nonumber
\end{eqnarray}

Note, in particular, that the eigenvalue $\lambda=1$ of
$\mathbf{Q}^M$ corresponds to the $M$ longitudinal eigenvalues
$\gamma_k=\exp(2\pi ik/M)$ of $\mathbf{Q}$, which are related to
directions longitudinal to the group synchronization manifold. Their
existence can already be seen solely by looking at $\mathbf{Q}$
itself, because its eigenvectors
\begin{equation}
  \label{eq:4}
  \mathbf{v}_k = \begin{pmatrix}
    \left.\begin{array}{c}
        \exp(-2\pi i k/M)\\
        \vdots\\
        \exp(-2\pi i k/M)
      \end{array}\right\}
    N_1\\
    \left.\begin{array}{c}
        \exp(-4\pi i k/M)\\
        \vdots\\
        \exp(-4\pi i k/M)
      \end{array}\right\}
    N_{2}\\
    \vdots\\
    \left.\begin{array}{c}
        \exp(-2\pi i k)\\
        \vdots\\
        \exp(-2\pi i k)
      \end{array}\right\}
    N_{M}
  \end{pmatrix},
\end{equation}
where each $\mathbf{v}_k$ corresponds to the longitudinal eigenvalue
$\gamma_k=\exp(2\pi ik/M)$, do not depend on the inner structure of
the blocks $\mathbf{A}^{(k)}$.

Given that the MSF is invariant with respect to rotations $\gamma
\rightarrow \exp(2\pi i/M)\gamma$ and that each of the multiple roots
of $\lambda_1,\ldots,\lambda_M$ are also rotations by multiples of
$2\pi/M$ with respect to the roots
$\{\sqrt[M]{|\lambda_1|}\exp[i\arg(\lambda_1)/M]$,$\ldots$,
$\sqrt[M]{|\lambda_M|}\exp[i\arg(\lambda_M)/M]\}$,
we can restrict ourselves to evaluating the master stability function
at the location of the eigenvalues
\begin{eqnarray}
  \label{eq:5}
  &&\{ \sqrt[M]{|\lambda_1|}\exp[i\arg(\lambda_1)/M], \ldots\\
  && \quad\ldots,\sqrt[M]{|\lambda_M|}\exp[i\arg(\lambda_M)/M], 0\}, \nonumber
\end{eqnarray}
which lie all inside the angular sector $\arg(\gamma) \in [0,2\pi/M)$. Note
that the zero eigenvalue is only added here if $n_0>0$, i.e., if at least
one block $\mathbf{A}^{(k)}$ differs from the others in size.

%

\end{document}